# Calculate electronic excited states using neural networks with effective core potential

JinDe Liu[1], Chenglong Qin[1], Xi He[1], and Gang Jiang[1†]

[1] Institute of Atomic and Molecular Physics, Sichuan University, Chengdu 610065, China

## Abstract

The crux of atomic structure theory, quantum chemistry, and computational materials science lies in determining the solutions to the multi-electron stationary Schrödinger equation. The neural network wave function method based on Quantum Monte Carlo (QMC) has achieved computational results of higher precision than existing post-Hartree-Fock methods across various systems, owing to the powerful expressive capability of the wave function itself and related outstanding optimization algorithms. However, the relative energy uncertainty of this method is limited to 0.01%, which presents challenges in solving problems like excited states and ionization energies, particularly in providing effective vertical excitation and ionization energies for elements beyond the fourth period. By considering effective core potentials to shield inner electrons, we can obtain higher precision vertical excitation and ionization energies, and their corresponding analytic wave functions, thus enabling more accurate calculations of physical quantities like transition dipole moments. The effectiveness of the effective core potential method is demonstrated through the calculation of ground state energies of elements from Lithium to Gallium, as the precision varies with the absolute energy of the system under study. Further, we calculated several energy levels and wave functions of atoms and molecules containing second and fourth period elements under a fixed spin configuration, illustrating the effectiveness of effective core potentials in the neural network wave function framework for multi-electron atomic and molecular excited state calculations. Utilizing the effective core potential method in conjunction with Ferminet, we achieved multiple excited state calculations in various atomic and molecular systems, for the first time achieving precision comparable to experimental data. This extends the method to levels of practical application value and establishes a new benchmark for ab initio theoretical excited state calculations.

## Introduction

The solution to the multi-electron stationary Schrödinger equation is a core issue in atomic structure theory, quantum chemistry, and computational materials science. It can reveal the comprehensive characteristics of a given atomic, molecular, or solid-state system, as well as the intrinsic mechanisms of various physical, chemical, and biological processes. [1–4].

However, solving this equation is extremely challenging and precise solutions are only possible for a limited number of toy models, such as the infinite potential well, harmonic oscillator, hydrogen-like atom, and Morse potential. As a result, a range of numerical methods have been developed, primarily categorized into two types: ground state and excited state solutions. Traditional ground state methods include Hartree-Fock (HF), Density Functional Theory (DFT), and post-Hartree-Fock approaches. These generally employ analytic integration and self-consistent iterations, but require stringent formal requirements for the wave function, necessitating the introduction of numerous assumptions. Although accuracy can systematically be improved through multi-configuration approaches, computational complexity increases significantly with enhanced precision. Traditional excited state methods include Time-Dependent DFT (TDDFT), Full Configuration Interaction (FCI), and Coupled Cluster (CC) methods, whose computational efficiency or accuracy are also greatly limited.

For instance, Time-Dependent Density Functional Theory (TDDFT) lacks accuracy and cannot provide the wave function [5–9]. Full Configuration Interaction (FCI) and Coupled Cluster (CC) methods offer high precision, but their computational efficiency is greatly limited [10,11]. Quantum Monte Carlo (QMC), which includes Variational Monte Carlo (VMC) and Fixed-Node Diffusion Monte Carlo (DMC), faces its own challenges. VMC generally produces energy with lower accuracy compared to DMC, so the typical QMC process involves using VMC to obtain a trial wave function and then employing DMC to calculate the final energy based on this function. However, the accuracy of DMC results is significantly limited due to the inability to modify the nodal surface of the trial wave function [10]. The subsequent invention of FCI-QMC greatly improved the nodal surface of trial wave functions and solution accuracy [12,13]. Additionally, the DMC method is limited to providing only energy values, which restricts calculations of other systemic properties [14]. Therefore, there is an urgent need for a method that can better address these challenges.

In recent years, with the advancement of deep learning technology, a method based on neural network wave functions has been developed, offering a novel approach and tool for solving the multi-electron stationary Schrödinger equation. This method, an unsupervised learning algorithm, utilizes neural networks as an approximation for the wave function. It then minimizes a suitable loss function through advanced optimization algorithms like Adam or KFAC, gradually ensuring that the proposed wave function meets physical constraints (including boundary conditions) and equations.

The neural network wave function method has several advantages: (1) Neural networks have a robust expressive capability, allowing them to fit wave functions of any complexity without the need for excessive assumptions. (2) It can simultaneously solve for both ground and excited states, along with corresponding wave functions and other physical quantities. (3) The method offers flexibility in choosing different Hamiltonians, variational forms, and electron coordinate sampling algorithms to cater to various problems and needs.

Key research areas in this method include the construction of trial wave functions, consideration of different Hamiltonians for various physical effects, choices of variational forms, design and selection of electron coordinate sampling algorithms, and exploration of principles for solving excited states. The trial wave functions generally consider electron-electron correlation (typically through electron cusps and Jastrow factors), optimization of wave function nodes (backflow functions, backflow displacement), and exchange antisymmetry (Slater determinants, Pfaffian determinants [15,16], Vandermonde determinants [17,18]). Prominent wave function models include FermiNet and its variants [19,20], PauliNet [4,21,22], PsiFormer [23,24], DeeperWin and its improvements [19,25], and DeepWF [17]. Equations include the non-relativistic full-electron Schrödinger equation, the relativistic full-electron Dirac equation, and the Schrödinger equation under effective core potentials [26]. The variational loss functions mainly comprise average energy [27–29], energy variance [30], and equation residuals [31–36]. Electron coordinate sampling algorithms include MCMC sampling and batch sampling. The core principle of solving for excited states is ensuring the orthogonality of wave functions across energy levels, mainly through minimizing overlap coefficients [4,29], constructing Schmidt-orthogonalized trial wave functions [37], and fixing the nodes of trial wave functions.

The neural network wave function method has achieved computational results of higher precision than existing post-Hartree-Fock methods in many systems, such as simple model explorations [29,30], calculations of ground [20,38,39] and excited states [4] of real atomic and molecular systems, efficient construction of molecular potential energy surfaces [23,40–43], and calculations of non-local potentials, including free electron gas [44] and real solid systems [45].

However, the neural network wave function method has some limitations. Primarily, the precision of the solution is directly related to the absolute energy of the system under study, with the final relative energy uncertainty only reaching about 0.01%. This means that when calculating systems with higher absolute energies, the final convergence energy uncertainty can be significant, making it challenging to obtain effective ionization energies or vertical excitation energies. To address this issue, one approach is to use effective core potentials (ECPs) to shield inner electrons, thereby significantly reducing the absolute energy of the system. Effective core potentials represent the interactions of atomic nuclei and inner electrons with an equivalent potential energy, simplifying the solution of multi-electron systems while preserving the characteristics of the outer electrons. By employing effective core potentials, we can enhance the precision and efficiency of the neural network wave function method without changing the relative energy uncertainty. This approach allows for the calculation of more accurate vertical excitation energies and ionization energies, along with their corresponding analytic wave functions, leading to more precise determinations of physical quantities like transition dipole moments.

We validated the accuracy of the effective core potential (ECP) method by calculating the ground state energies of elements from Lithium (Li) to Gallium (Ga), demonstrating its efficacy through the variation of absolute energy in the systems under study. Furthermore, we computed the energies and wave functions of several energy levels of atoms and molecules containing elements from the second and fourth periods, under fixed spin configurations. This showcased the effectiveness of effective core potentials in the neural network wave function framework for solving excited states of multi-electron atomic and molecular systems.Additionally, we calculated several excited states of the benzene molecule with high precision. By integrating the effective core potential method with FermiNet, we achieved multiple excited state calculations in various atomic and molecular systems, attaining a level of accuracy comparable to experimental data for the first time. This advancement extends similar methods to a level of practical application value, establishing a new benchmark for ab initio theoretical calculations of excited states.

## Results

### Algorithmic framework

The non-relativistic Schrödinger equation for a molecule, within the Born–Oppenheimer approximation and defined by a Hamiltonian $\hat{H}$ that incorporates the charges Z and coordinates ($R_I$) of the nuclei, is a second-order differential equation. This equation determines the wavefunction, $\psi_n(\vec{r}_1,...,\vec{r}_m)$, which depends on the coordinates of N electrons.

$$\hat{H}\psi_n(\vec{r}_1,...,\vec{r}_m) = E\psi_n(\vec{r}_1,...,\vec{r}_m) \qquad (1)$$

$$\hat{H} = -\sum_i \frac{1}{2}\nabla_i^2 + \sum_{i<j}\frac{1}{|\vec{r}_i-\vec{r}_j|} - \sum_{i,I}\frac{Z_I}{|\vec{r}_i-\vec{R}_I|} + \sum_{I<J}\frac{Z_I Z_J}{|\vec{R}_I-\vec{R}_J|} \qquad (2)$$

For any energy level n, its wavefunction $\psi_n(\vec{r}_1,...,\vec{r}_m)$ is orthogonal to the wavefunctions $\psi_i(\vec{r}_1,...,\vec{r}_m)$ of all energy levels i less than or equal to n, denoted as $\langle\psi_n|\psi_i\rangle = 0, (i=1,2,...,n-1)$. An alternative approach to the Schrödinger equation involves the concept of an expectation value. Rather than directly solving equation (1), the solution for any state can be obtained by minimizing the energy expectation value across all possible wavefunctions, as per the variational principle.

$$E_n = \min_{\psi_n}\langle\psi_n|\hat{H}|\psi_n\rangle \qquad (3)$$

Using the neural network wave function FermiNet as the trial wave function $\psi_n^\theta$, its framework is depicted in Figure 1 (a). To ensure the orthogonality of the wave function of the energy level being solved with those of known energy levels, we incorporate the overlap coefficients $S_{ni}^\theta = \langle\psi_n^\theta|\psi_i\rangle (i=1,2,...,n-1)$ as minimization terms into the variational loss function. With the average energy denoted as $E_n^\theta = \langle\psi_n^\theta|\hat{H}|\psi_n^\theta\rangle$, the total variational loss function is then

established as:

$$L_n(\theta) = E_n^\theta + \alpha \sum_{i<n} S_{ni}^\theta. \tag{4}$$

In the description above, $\theta$ refers to the parameters to be solved for within the trial wave function.

In the context of neural network wave functions, the above integrals require Monte Carlo integration. Therefore, within this framework,

$$S_{ij}^\theta = \sqrt{\mathbb{E}_i\left[\frac{\varphi_j(r_i)}{\varphi_i(r_i)}\right]\mathbb{E}_j\left[\frac{\varphi_i(r_j)}{\varphi_j(r_j)}\right]} \tag{5}$$

$$E_n^\theta = \mathbb{E}_n\left[E_{loc}^\theta(\tilde{r}_n)\right] \tag{6}$$

Where $\tilde{r}_n$ represents the coordinates of all electrons at energy level n, and $E_{loc}^\theta(\tilde{r}_n)$ is the local electron energy, under the full-electron Hamiltonian we have:

$$E_{loc}^\theta(\tilde{r}) = \frac{\hat{H}\psi_n^\theta(\tilde{r})}{\psi_n^\theta(\tilde{r})} = \frac{\hat{T}\psi_n^\theta(\tilde{r})}{\psi_n^\theta(\tilde{r})} + V(\tilde{r}) \tag{7}$$

Where $\hat{T} = -\sum_i \frac{1}{2}\nabla_i^2$ represents the kinetic energy operator. To facilitate the calculation of second derivatives of the determinant wave functions, its equivalent form is used to compute the local kinetic energy:

$$\frac{\hat{T}\psi_n^\theta(\tilde{r})}{\psi_n^\theta(\tilde{r})} = -\frac{1}{2}\sum_d\left[\frac{\partial^2 \log|\psi|}{\partial x_d^2} + \left(\frac{\partial \log|\psi|}{\partial x_d}\right)^2\right] \tag{8}$$

$V(\tilde{r})$ represents the local potential energy, which under the full-electron context is defined as

$$V(\tilde{r}, \tilde{R}) = \sum_{i<j}\frac{1}{|\vec{r}_i - \vec{r}_j|} - \sum_{i,I}\frac{Z_I}{|\vec{r}_i - \vec{R}_I|} + \sum_{I<J}\frac{Z_I Z_J}{|\vec{R}_I - \vec{R}_J|}. \tag{9}$$

$Z^{eff} = Z - Z^{core}$ denotes the effective nuclear charge number, and $Z^{core}$ represents the number of core electrons. Therefore, under the effective core potential, the local potential energy is defined as

$$V(\tilde{r}, \tilde{R}) = \sum_{i<j}\frac{1}{|\vec{r}_i - \vec{r}_j|} - \sum_{i,I}\frac{Z_I^{eff}}{|\vec{r}_i - \vec{R}_I|} + \sum_{I<J}\frac{Z_I^{eff} Z_J^{eff}}{|\vec{R}_I - \vec{R}_J|} + V_{ECP}(\tilde{r}, \tilde{R}), \tag{10}$$

where $V_{ECP}(\tilde{r}, \tilde{R})$ represents the effective core potential, which is given by

$$V_{ECP}(\vec{r}) = \sum_{v=1}^{n_v} V_{loc}(r_v) + \sum_{vlm} V_l(r_v) Y_{lm}(\Omega_v) \int d\Omega_v Y_{lm}^*(\Omega_v') \frac{\psi_n^\theta(r_1,...,r_v',...r_N)}{\psi_n^\theta(r_1,...,r_v,...r_N)}. \quad (11)$$

Where $V_{loc}(\vec{r}) = \sum_k A_{lk} e^{-B_{lk}r^2}$ is the local effective core potential, and the term containing $V_l(\vec{r}) = r^{-2} \sum_k A_{lk} r^{n_{lk}} e^{-B_{lk}r^2}$ represents the non-local effective core potential, with $A_{lk}, B_{lk}$ being the known parameters of the effective core potential.

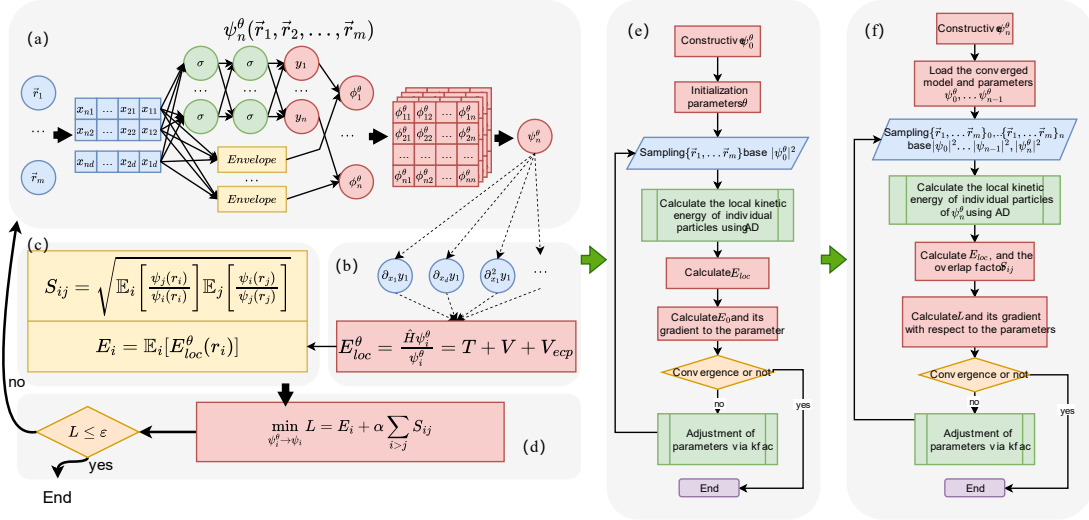

Figure 1 Algorithmic framework of the multi-electron fixed-state Schrödinger equation multi-excited-state solver, (a) Fitting the wavefunction for the neural network (b) Using automatic differentiation to find the Hamiltonian quantity of the system at each sampling point, (c) Calculating the individual loss functions using Monte Carlo integration, (d) Solving the equation by minimizing the total loss function. (e) Base state solution process, (f) excited state solution process.

The overall algorithm framework, as shown in Figures 1 (a)-(d), begins with mapping electron coordinates into single-electron and two-electron features through symmetry transformations. These features are then combined into electron orbitals using neural networks or certain parametric functions. Subsequently, these orbitals form determinant wave functions. Local energies are derived using automatic differentiation techniques, which further facilitate the calculation of the total loss function. This loss function is then minimized step by step to optimize the wave function and energy. Specific procedures for solving ground and excited states are depicted in Figures 1 (e) and (f) respectively. The primary distinction between the two lies in the fact that solving for excited states requires multiple samplings of all wave functions to compute overlap coefficients. Additionally, the optimization process necessitates the simultaneous minimization of these overlap coefficients. This is a method that gets multiple states by multiple times training. The basic procedure of the method is as follows:

1) The parameters resulting from the iteratively solved wave function are immobilized to obtain wave functions $\varphi_i (i=1,2,...,n-1)$, then parameterized wave functions $\varphi_n^\theta$ are designated for subsequent solving.

2) We minimize the following loss function to solve the wave function.

$$L = \mathbb{E}_n \left[ E_{loc}^\theta (\tilde{r}_n) \right] + \alpha \sum_{i \neq n} \mathbb{E}_i \left[ \frac{\varphi_n(r_i)}{\varphi_i(r_i)} \right] \mathbb{E}_n \left[ \frac{\varphi_i(r_n)}{\varphi_n(r_n)} \right]$$

3) We freeze the parameters of the solved wave function to obtain wave functions $\varphi_n$, and then incorporate wave functions $\varphi_i (i=1,2,...,n)$ to obtain wave functions $\varphi_i (i=1,2,...,n+1)$.

By repeating steps from 1 to 3, we can achieve multiple iterations of training to solve multiple energy levels.

## Calculate excited states for atoms

We optimized the lowest several excited states of some atoms and calculated their vertical excitation energies (see Supplementary Table I), with each wave function comprising 16 determinants. In all systems, we achieved high-precision total energies and estimated energies for the first few excited states, comparable to high-precision quantum chemistry methods. For atoms like Li, Be, B, Ga, Br, we calculated only one spin configuration, while for other atoms, we computed two spin configurations, as detailed on the x-axis of Figure 4(a). The vertical excitation energies are shown in Figure 4(a), and their convergence curves are presented in FIG S4.

Compared to the NIST's optimal reference values, we obtained errors within 10 mHa for almost all energy levels, a significant improvement over the 34 mHa error of PauliNet. Notably, our results for the N atom were distinctly better than those of PauliNet. The O atom, having the highest number of outer electrons, showed the greatest uncertainty within the same number of iterations, reaching about 10 mHa. The third excited state of the Ga atom exhibited a split in the convergence process, likely due to errors inherent in the ccecp pseudopotential. Under the (3,0) spin configuration for the As atom, the third energy level displayed eight degenerate states. Notably, it is challenging to distinguish the energies of $4s^24p^2(^3P)5p\ 4D$ and $4s^24p^2(^3P)5p\ ^4P$ which are within error margins. These eight degenerate states correspond to these two energy levels, satisfying the requirements for degeneracy. It's important to note that our method does not consider spin-orbit coupling and thus cannot differentiate energy levels of the same spectral term with different J values.

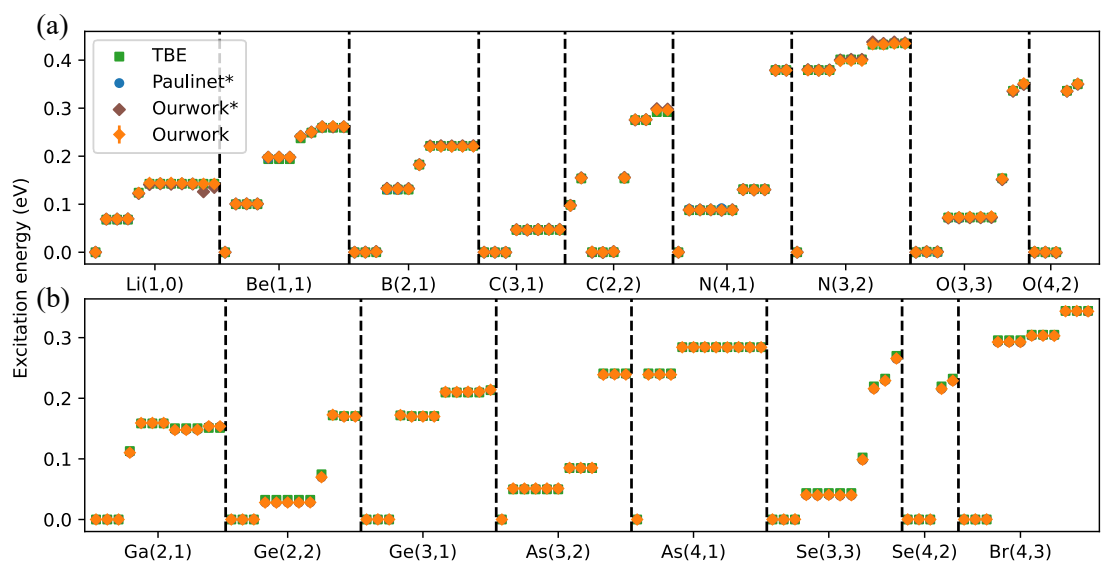

FIG 2 High precision vertical excitation energy of single atom. (a) The results of Li-N calculation with ccecp[He] (green diamond) used Ferminet are compared with the theoretical best estimate (TBE) in the NIST database (orange square) and calculation (purple circle) with all-electrons used PauliNet. (b) The Ga-Br calculated by Ferminet with ccecp[Zn] was compared with the theoretical best estimate (TBE) in the NIST database (orange square).

The error analysis of some atoms compared with PauliNet is shown in Figure 5. It's evident that across numerous wave functions and five to six energy levels, our Mean Absolute Error (MAE) is within 3 mHa. Compared to PauliNet, on atoms like Li and Be, our MAE is significantly larger, partly due to the greater number of energy levels we solved for. Another factor might be the ineffectiveness of the effective core approximation for the excited states of Li/Be. Notably, even with more energy levels, we still achieved smaller MAE in larger systems like N and O. It's also worth mentioning that variance matching did not significantly improve the convergence results in atomic systems. On the Li atom, variance matching actually worsened our results, likely because the ground state variance converges much faster than that of excited states. Thus, some excited states, when variance matched to a not-yet-correctly-converged ground state energy, led to larger errors. The convergence curve for this is seen in the supplementary FIG. S5(b). It's also readily observable that with the full-electron PauliNet calculations, the error steadily increases as the atomic number grows. In contrast, with Ferminet-ecp, the error remains almost unchanged, showing no significant increase with rising atomic numbers. The error analysis for the other atoms is available in the supplementary FIG. S4.

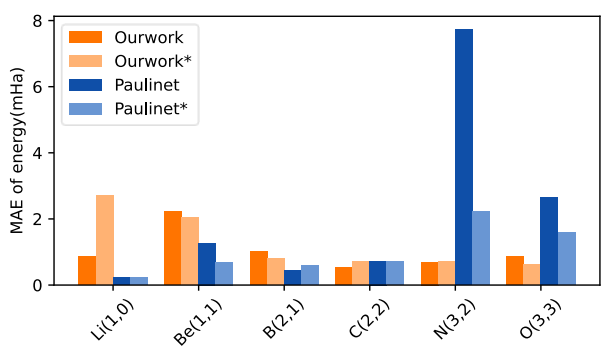

FIG 3 Mean absolute errors (MAE) of vertical excitation energies compared with the best theoretical estimates for Li-N，and the * sign indicates the result of variance matching. (See Supplementary Table II for numerical data; source data are provided as data files).

The high symmetry of atoms leads to multiple orthogonal states with the same energy, known as degenerate states. To study this phenomenon, we further analyzed the degeneracy of the results shown in Figure 4(a), as illustrated in Table 1. Due to the combined constraints of orthogonality and energy minimization, our method can achieve convergence from lower to higher energy levels, naturally resulting in multiple orthogonal states for degenerate energy levels. This can then be compared with the multiplicity of each energy level derived from theoretical atomic electron configurations. We found that our method aligns with theoretical degeneracies on most energy levels, with some deviations only at the highest energy levels (as indicated in green in Table 1), likely due to an insufficient number of wave functions calculated. Additionally, we identified a special case in the Ge atom's $4s^24p^25s$ $^3P$ energy level, which displayed four identical energies instead of the theoretically expected threefold degeneracy. An analysis of the overlap matrix revealed that this was due to suboptimal convergence under the orthogonality condition for that energy level. Notably, our method performed well in calculating the excited states of lower energy levels (the first six levels), without failing to converge to the target excited states, a problem PauliNet encountered with the 3rd excited state of Be. Our approach is suitable for computing multiple degenerate states of atoms, but attention must be paid to the convergence of orthogonality conditions and the completeness of the wave functions to be solved.

Table 1 Analysis of the recovery of the atomic simplicity. The data in the table are the simplicity of a particular spectral term, which corresponds to the total orbital quantum numbers L of 0,1,2 corresponding to S,P,D, and the theoretical simplicity of 1, 3, and 5, respectively.

|   | Li | Be | B | C | N | O | Ga | Ge | As | Se | Br |
|---|---|---|---|---|---|---|---|---|---|---|---|
| S | 1,1 | 1,1,1 | 1 | 1,1,1 | 1,1 | 1,1,1,1 | 1 | 1 | 1,1 | 1,1,1,1 |  |
| P | 3,3 | 3,3 | 3,3 | 3,3,2 | 3,2,3,3 | 3,3 | 3,3 | 3,4,3,3 | 3,3,3,3 | 3,3,1 | 3,3,3,3 |
| D | 4 | 3 | 5 | 5,2 | 5,4 | 5 | 5 | 5,5 | 5,5 | 5 |  |

FermiNet uses two determinants to represent electrons of different spins, which implies that the final spin multiplicity is mixed. Taking the carbon atom (C) as an example, with valence electrons in the 2p² configuration, the four common eigenstates of the normalized $(S^2, S_z)$ [using the $(s_{1z}, s_{2z})$ representation] can be represented as:

$$\chi_{1,1} = |\uparrow\uparrow\rangle_{12}$$
$$\chi_{1,-1} = |\downarrow\downarrow\rangle_{12}$$
$$\chi_{1,0} = \frac{1}{\sqrt{2}}\left[|\uparrow\downarrow\rangle_{1,2} + |\downarrow\uparrow\rangle_{12}\right] \quad (12)$$
$$\chi_{0,0} = \frac{1}{\sqrt{2}}\left[|\uparrow\downarrow\rangle_{1,2} - |\downarrow\uparrow\rangle_{12}\right]$$

It's clear that the spin state of the carbon atom (C) with a spin configuration of (3,3) is a linear combination of $|\uparrow\downarrow\rangle_{1,2}, |\downarrow\uparrow\rangle_{12}$ spin states, consisting of singlet state $\chi_{0,0}$ and triplet states $\chi_{1,0}$. For the C atom with a spin configuration of (4,2), the spin states of the valence electrons are represented by $|\uparrow\uparrow\rangle_{1,2}, |\downarrow\downarrow\rangle_{12}$, and the possible eigenstates include only triplet states $\chi_{1,1}$ and $\chi_{1,-1}$. This is consistent with the energy levels of the C atom observed in our experiments.

Further analysis of the orbitals constituting the ground state wave function of the C atom reveals clear features of the 2s and 2p orbitals. This is not a result of the trial wave function but rather a requirement of the Schrödinger equation for the C atom itself. This underscores the fundamental nature of the atomic structure as dictated by quantum mechanics, independent of the specifics of the wave function modeling approach.

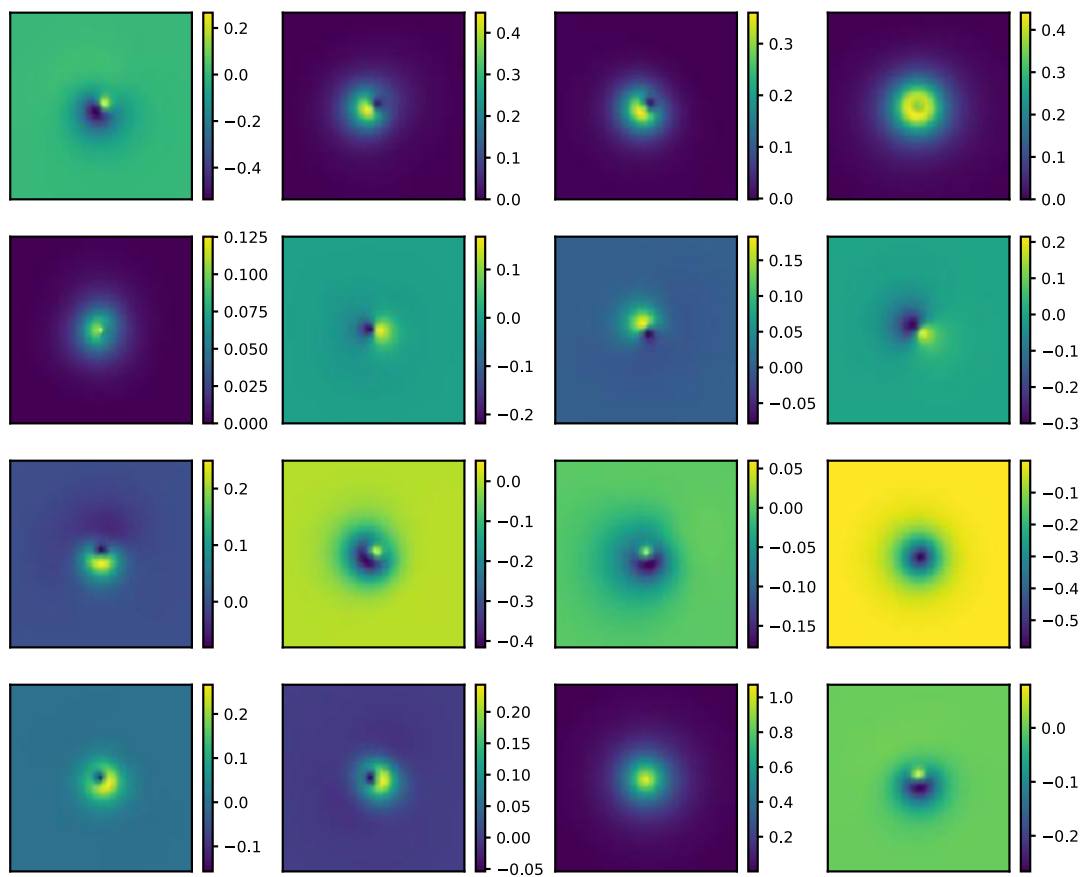

Figure 4 Slice diagram of the wave function of some of the orbitals of a C atom. It is plotted by fixing the coordinates of the other electrons to 0 and moving only one electron in the [-3,3]Bohr range of the x-y plane to calculate the corresponding values for each orbit.

## Calculate excited-state for middle and large molecular results

We calculated the ground state energies of molecules with fixed geometries, and the results are shown in Figure 2(c). Compared to the high-precision method CCSD(T)/CBS, only the CO system had an energy 10 mHa higher in our results, but it was still very consistent with the high-precision calculations of CCSD(T) with ccecpccpv5z. For the H2CSi system, our results closely matched those of CCSD(T)/CBS. In other systems, we achieved results several mHa lower than CCSD(T)/CBS, demonstrating the effectiveness of neural network wave functions with effective core potentials in molecular ground state calculations.

Further, we calculated the vertical excitation energies of more molecules with fixed geometries, as shown in Figure 2(a), with specific data in Supplementary Table S2. In each experimental round, we sequentially optimized dozens of excited states and initialized all parameters of the neural network wave function using Xavier initialization. As these molecules satisfy certain symmetries, their excited states also exhibit degeneracy, which we accounted for. We identified different excitation energies for 10 (LiH), 9 (BeH), 7 (CO), 6 ($H_2O$), and 8 ($H_2S$) states, matching the reference values and, in some systems, even aligning better with experimental values for certain energy levels. However, results like the 2nd excited state of BeH, the 4th of $H_2O$, and the 3rd and 6th of $H_2S$ were consistent with existing methods but deviated from experimental values.

Compared to PauliNet, our results showed a Mean Absolute Error (MAE) against TBE as depicted in Figure 2(b), with asterisks indicating variance-matched results. Notably, we achieved an MAE of less than 5 mHa in both cases, with most energy level errors under 1.5 mHa, reaching chemical accuracy. Without variance matching, only in the LiH system did we have a larger error than PauliNet, likely due to calculating 12 wave functions compared to their 8. However, the error was indeed more significant compared to other systems, possibly because the Li atom's ccecp pseudopotential considered only one free electron, insufficient for excited state calculations of Li. The relatively poor results for single-atom Li, Be, and BeH calculations corroborate this. Also, variance-matched results were significantly better than those without, consistent with PauliNet.

We also tried zero-variance extrapolation and convergence curve fitting extrapolation methods, as shown in Supplementary Figure S4. These methods depend heavily on data selection and often fail to yield satisfactory results, making them less practical for excited state problems. Compared to PauliNet, which can only stably obtain the first excited state, our method can stably reach up to the 6th excited state, which is undoubtedly very useful for studying the excited states of atomic and molecular systems. However, for closely spaced higher excited state clusters, we often couldn't achieve clear energy level separation, a common issue in high excited state calculations with general methods. Importantly, our method effectively determines molecular degeneracy, which is significant for analyzing molecular excited states, such as potential energy surface crossings.

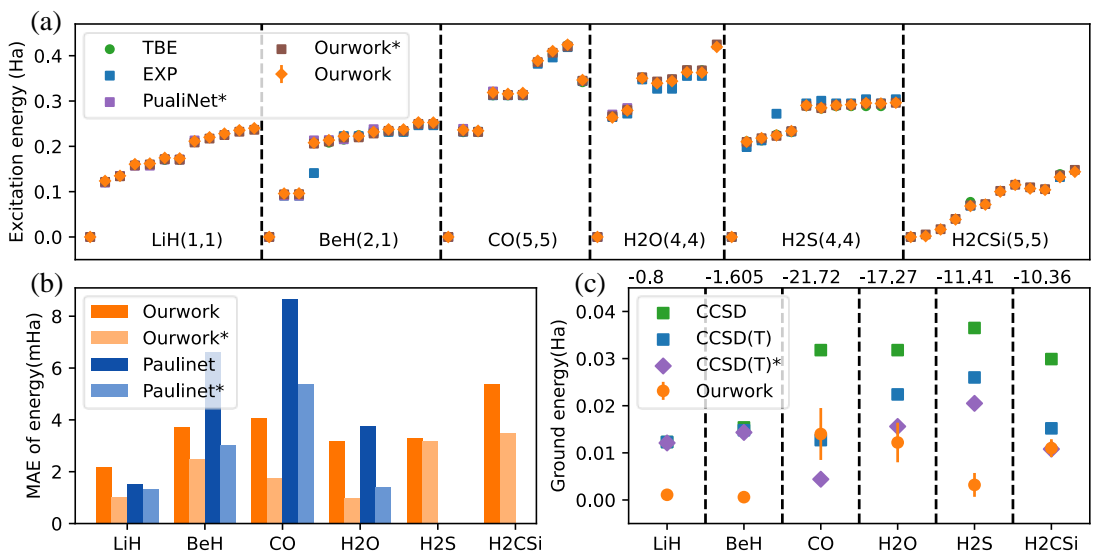

FIG.5 Deep VMC with ECP obtains highly accurate excited-state energies and wavefunctions for small molecules. (a) Vertical excitation energies of various small molecular systems, with geometric parameters given in Supplementary Information S4; (b) Mean absolute errors (MAE) of vertical excitation energies compared with the best theoretical estimates for each system;(c) Comparison of ground-state results with CC methods, where CCSD and CCSD(T) are calculated with the ccecpccpv5z basis set, and CCSD(T)* is obtained by extrapolating HF and CCSD(T) to the complete basis set (CBS) limit. Detailed data are provided in Supplementary Table S3.

Utilizing the ground state as an initial guess, we successfully applied this method to calculate the first few excited states of the benzene molecule. Given its significance in biology and organic chemistry, extensive research has been conducted on its electronic structure and other properties. However, a full-electron calculation of benzene, with its 42 electrons, demands a high level of accuracy in describing its electronic states, which can be challenging depending on the theoretical level used. Using FermiNet with only 16 determinants, we obtained very accurate total energies for both the ground state and the first excited state (as shown in the top left of Figure 7). Our results are comparable to those from CCSD(T)/ccecpccpvQZ, indicating a high precision of the wave function suitable for calculating other observables. The calculated excitation energies are also displayed (on the right side of Figure 4), with comparisons made between FermiNet with ecp, PauliNet, and several experimental and theoretical results. Our results were closer to experimental values than PauliNet's without variance matching, but the outcomes from variance matching and both extrapolation methods were less accurate than PauliNet. This discrepancy might be due to the initialization of the excited states using ground state calculations.

In Figures 4b and 4d, we display two-dimensional slices of the trial wave functions for the ground state and first excited state. These slices were generated by moving a single spin-up electron within a two-dimensional box, while all other electrons were fixed at representative positions proposed by Liu et al. [39,46], as shown in Figure 4a. Comparing Figure 4b (ground state wave function) with Figure 4d (excited state), we notice that the node patterns, represented by darker pixels, are somewhat similar for both energy levels. However, there are clear differences in the nodal topologies between Figure 7(b) for the ground state and Figure 7(d) for the excited state, particularly in the lower section marked in red. These differences are due to the orthogonality between the ground and excited state wave functions causing variations in the nodal surfaces. Red dots in the figures indicate electrons, and all are situated on the nodal planes, a result of the anti-symmetry constraint inherent in the wave function for electrons of the same spin.

Additionally, in Figures 4b and 4d, the lighter areas, representing regions where the wave function values are larger (nodes), are very close to each other, consistent with references [39,46]. We also present slices of the two-dimensional electron density for benzene, derived from the sum of the squares of the orbitals in our trial wave function (the orbitals of the first determinant as shown in FIG.S5). It's evident that electrons are primarily distributed around the nuclei, and the bonding in benzene is clearly visible (i.e., higher electron density between C-C and C-H compared to other areas). The electron density distributions for the ground state and excited state are similar.

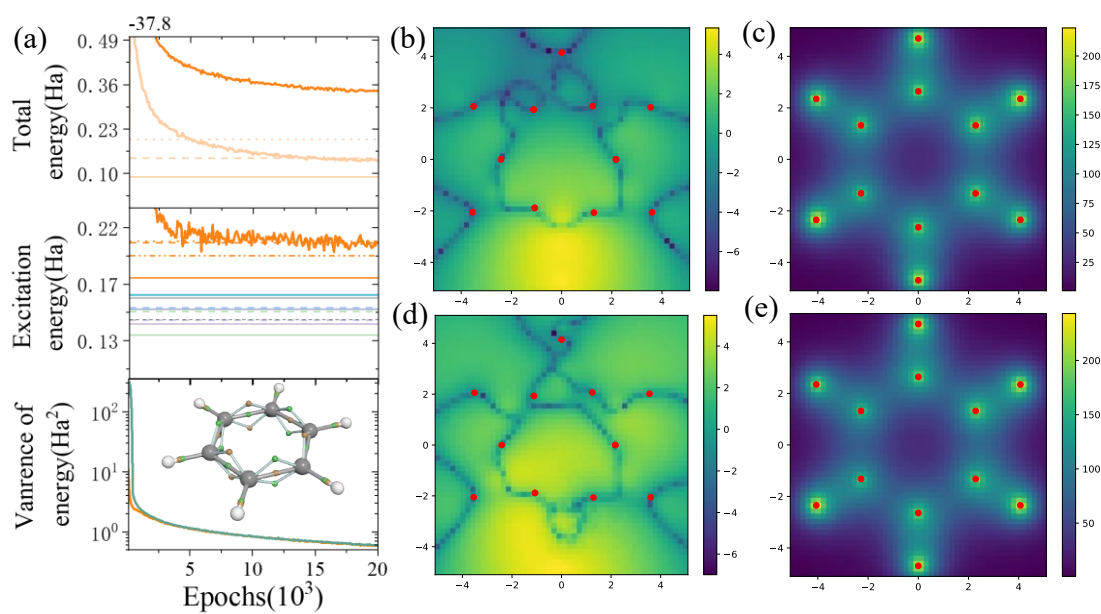

FIG.6 We performed calculations on the two lowest electronic states of the benzene molecule. In part a, the total energies and their corresponding variances for the ground state (in light orange) and excited state (in orange) are shown converging with training. The graph also displays the ground state total energy using the frozen-core approximation CCSD(T) with aug-cc-pVnZ basis sets (n = D, T) (represented by the blue dashed line), and the complete CCSD(T) total energy at the CBS limit (depicted by the blue solid line) [36]. Part c presents excitation energies calculated using various methodologies: PauliNet, TDDFT [68], CC [78], DMC [26], and CAS-PT (as implemented in openMolcas [80] and [79]), along with experimental values for excitation energies (both adiabatic, shown with a black dashed line, and vertical, depicted by a black solid line). Numerical data for these calculations are available in Supplementary Table IV, and the source data are provided in a source data file format.

## Discussion

The challenge of solving the multi-electron Schrödinger equation can be broken down into three dimensions: Scaling with Number of Electrons: As the number of electrons N increases, the dimensionality of the problem scales by 3N, and the computational complexity of various algorithms rises, ranging from $N^3$ to $N^8$. Influence of Atomic Number: With a constant number of electrons, an increase in atomic number accentuates electron localization and relativistic effects. This leads to a growing significance of electron-electron correlation effects, reducing precision and making convergence more difficult. Complexity with Rising Energy Levels: Within the same system, as energy levels rise, the number of wave function nodes and overall complexity increase, negatively impacting convergence. The effective core approximation can mitigate electron localization issues in heavy-element systems and decrease the number of electrons that need to be solved for.

In our numerical experiments on atomic systems, under full-electron conditions with consistent numbers of outer electrons, we observed that as the system's absolute energy decreases, the variance in energy that can be converged upon at 20k Epochs progressively worsens. This is attributed to increased electron localization as the atomic number rises, causing electron position sampling to be closer to the nuclei and amplifying numerical errors. Under effective core approximation, where core electrons are frozen, this effect is significantly reduced, enhancing the system's convergence. However, as systems grow larger, the magnitude of variance achievable also increases, indicating that effective core potentials do not fundamentally resolve the shortcomings caused by electron localization. Additionally, the variance increases with higher energy levels due to the rising complexity of the corresponding wave functions, reducing convergence under the same wave function parameter scale. Particularly, in O/N atoms, when energy levels exceed 7, severe instability in convergence or even failure is observed, suggesting a limit to the energy levels that can be resolved by neural network wave functions with finite parameters.

Furthermore, an error analysis of vertical excitation energies across all atoms revealed that, unlike PauliNet, which shows a significant increase in error with larger systems, our method maintains a mean absolute error within 5 mHa. A detailed analysis of the degeneracy of the obtained energy levels showed that we could reliably reproduce the theoretical degeneracies for atomic systems. Even in cases like the Ge atom, where degeneracy recovery failed at certain levels, judgments could be made based on the convergence of orthogonality conditions. This demonstrates the practical potential of FermiNet with ECP in actual atomic systems.

In molecular systems, our results consistently outperform those provided by PauliNet. In larger systems like H2S and H2CSi, we achieved a mean absolute error of only 5 mHa, reaffirming the method's advantage of not exhibiting significant error increases with larger systems. Our experiments with benzene further validate the method's effectiveness in molecules composed of light elements, with the advantage still being more pronounced in heavy-element atoms.

## Methods

### Neural network wave function

Neural network wave functions typically comprise four main components: Feature Input Layer: This layer includes the relative position vectors between electrons and nuclei (e-n), the distances between electrons and nuclei (ren), electron-electron (e-e) position vectors, and electron-electron distances (ree). It constructs electron-nuclei (e-n) features, encompassing both original and scaled features. Equivariant Mapping Layer: This layer consists of single-electron feature mapping and two-electron feature mapping (with spin partitioning). The purpose of this layer is to ensure that the mapping preserves the physical symmetries of the system. Orbital Mapping Layer: This layer includes mappings for single-electron and multi-electron orbitals. It translates the features from the previous layer into a form suitable for constructing the wave function. Determinant Wave Function Mapping Layer: The primary component here is the Slater determinant. This layer constructs the final wave function by combining the single and multi-electron orbital mappings into determinant form, which is essential for ensuring the anti-symmetry of the wave function with respect to electron exchange.

In FermiNet, the trial wave function orbitals $\phi_n^\theta(\vec{r}) = N_n^\theta(\vec{r}) \cdot F_n^\theta(\vec{r})$, as depicted in Figure 1(a), are based on a multi-input, multi-output neural network. The output node $N_n^\theta(\vec{r})$ of this network is from a fully connected layer that processes multiple inputs and outputs. The input layer consists of single-electron and multi-electron features, which are mappings of the coordinates of electrons and nuclei [20,24].

Specifically, the primary function $N_n^\theta(\vec{r})$ of $\phi_n(\vec{r})$ is to learn and fit the details of the wave function orbitals, leveraging the universal approximation theorem of neural networks [47]. This theorem allows the neural network to approximate a wide range of functions, making it highly effective for modeling complex wave functions in quantum systems. Additionally, $F_n^\theta(\vec{r})$ is an envelope function designed to ensure that the trial wave function satisfies boundary conditions. This aspect of the network is crucial for maintaining physical realism in the modeled system, as it ensures the wave function behaves correctly at the limits of the space being considered. The overall trial wave function is given by equation (13) in references [20,26].

$$\varphi_\theta(\vec{r}) = \exp\left(\gamma(\vec{r}) + J_\theta\left(x(\vec{r};\vec{R})\right)\right) \sum_p c_p \det\left[\phi_{\theta,d}^\uparrow\left(\vec{r}, x(\vec{r};\vec{R})\right)\right] \det\left[\phi_{\theta,d}^\uparrow\left(\vec{r}, x(\vec{r};\vec{R})\right)\right] \quad (13)$$

Where $\gamma(\vec{r}) = \sum_{i<j} \frac{r_{ij}}{1+r_{ij}}$ is electron cusp condition, $J_\theta\left(x(\vec{r};\vec{R})\right)$ is Jastrow factor, $\phi_{\theta,d}^\uparrow\left(\vec{r}, x(\vec{r};\vec{R})\right)$ is the orbital wave function.

Due to the complexity of the trial wave function and the requirements of stochastic gradient descent, the calculation of various loss functions necessitates the use of automatic differentiation techniques [48] and Monte Carlo integration [49]. Automatic differentiation enables the efficient and accurate computation of derivatives, which are essential for gradient-based optimization methods like stochastic gradient descent. This technique is particularly valuable in the context of complex neural network architectures, where manual differentiation would be prohibitively cumbersome and error-prone. Monte Carlo integration is employed to handle the integrals that arise in quantum mechanical calculations, especially when dealing with high-dimensional spaces typical in many-electron systems. This method is well-suited for calculating expectations and energy values, which are integral to loss functions in quantum chemistry models. Monte Carlo integration's ability to handle complex, multi-dimensional integrals makes it a key tool in the computational toolkit for implementing neural network-based quantum chemistry methods.

## Mean value and gradient calculation of operator

To facilitate the computation of loss functions, their gradients, and other physical quantities within the Quantum Monte Carlo (QMC) framework, we consider the general expression for the expectation value of a Hermitian operator, as given in equation (14) [4]. The calculation of such expectation values is essential for determining the properties of quantum systems.

$$\hat{O}_{ij} = \langle \psi_i^\theta | \hat{O} | \psi_j^\theta \rangle = \text{sgn}\left( \mathbb{E}_i \left[ \frac{\hat{O}\psi_j^\theta(r)}{\psi_i^\theta(r)} \right] \right) \times \sqrt{ \mathbb{E}_i \left[ \frac{\hat{O}\psi_j^\theta(r)}{\psi_i^\theta(r)} \right] \mathbb{E}_j \left[ \frac{\hat{O}\psi_i^\theta(r)}{\psi_j^\theta(r)} \right] } \quad (14)$$

This expression simplifies to pairwise overlaps (see Eq.(14)) when setting $\hat{O} = \hat{I}$. The gradient of the operator with respect to the undetermined parameters of the wave function is given by:

$$\partial \hat{O}_{ij} = \frac{1}{\hat{O}_{ij}} \left\{ \begin{array}{l} \mathbb{E}_i \left[ \left( \frac{\hat{O}\psi_j^\theta(r)}{\psi_i^\theta(r)} - \mathbb{E}_i \left[ \frac{\hat{O}\psi_j^\theta(r)}{\psi_i^\theta(r)} \right] \right) \partial \ln |\psi_i^\theta(r)| \right] \times \mathbb{E}_j \left[ \frac{\hat{O}\psi_i^\theta(r)}{\psi_j^\theta(r)} \right] \\ + \mathbb{E}_j \left[ \left( \frac{\hat{O}\psi_i^\theta(r)}{\psi_j^\theta(r)} - \mathbb{E}_i \left[ \frac{\hat{O}\psi_i^\theta(r)}{\psi_j^\theta(r)} \right] \right) \partial \ln |\psi_j^\theta(r)| \right] \times \mathbb{E}_i \left[ \frac{\hat{O}\psi_j^\theta(r)}{\psi_i^\theta(r)} \right] \end{array} \right\} \quad (15)$$

The expression for the expectation value of a single wave function with respect to a Hermitian operator, and the gradient of this operator with respect to the adjustable parameters of the wave function, are easily derived as shown in equations (16) and (17).

$$\hat{O}_i = \langle \psi_i^\theta | \hat{O} | \psi_i^\theta \rangle = \mathbb{E}_i \left[ \frac{\hat{O}\psi_i^\theta(r)}{\psi_i^\theta(r)} \right] \quad (16)$$

$$\partial \hat{O}_i = 2\mathbb{E}_i \left[ \left( \frac{\hat{O}\psi_i^\theta(r)}{\psi_i^\theta(r)} - \mathbb{E}_i \left[ \frac{\hat{O}\psi_i^\theta(r)}{\psi_i^\theta(r)} \right] \right) \partial \ln |\psi_i^\theta(r)| \right] \quad (17)$$

This expression simplifies to the average energy, as shown in Eq.(6), upon setting $\hat{O} = \hat{H} = \hat{T} + \hat{V} + \hat{V}_{ECP}$. Consequently, the gradients for the average energy and the overlap coefficients are easily obtained.

$$\partial S_{ij} = \frac{1}{S_{ij}} \left\{ \begin{array}{l} \mathbb{E}_i\left[\left(\frac{\psi_j^\theta(r)}{\psi_i^\theta(r)} - \mathbb{E}_i\left[\frac{\psi_j^\theta(r)}{\psi_i^\theta(r)}\right]\right)\partial \ln|\psi_i^\theta(r)|\right] \times \mathbb{E}_j\left[\frac{\psi_i^\theta(r)}{\psi_j^\theta(r)}\right] \\ + \mathbb{E}_j\left[\left(\frac{\psi_i^\theta(r)}{\psi_j^\theta(r)} - \mathbb{E}_j\left[\frac{\psi_i^\theta(r)}{\psi_j^\theta(r)}\right]\right)\partial \ln|\psi_j^\theta(r)|\right] \times \mathbb{E}_i\left[\frac{\psi_j^\theta(r)}{\psi_i^\theta(r)}\right] \end{array} \right\} \quad (18)$$

$$\partial E_i = 2\mathbb{E}_i\left[\left(\frac{\hat{H}\psi_i^\theta(r)}{\psi_i^\theta(r)} - \mathbb{E}_i\left[\frac{\hat{H}\psi_i^\theta(r)}{\psi_i^\theta(r)}\right]\right)\partial \ln|\psi_i^\theta(r)|\right] \quad (19)$$

## Calculation of Leapfrog Parameters

The transition oscillator strength is known to be:

$$f_{ij} = \frac{2(E_i - E_j)}{3}\sum_k \left|\langle\psi_i^\theta|\vec{r}_k|\psi_i^\theta\rangle\right|^2. \quad (20)$$

Based on equation (13), the formula for calculating the oscillator strength can be easily derived as:

$$f_{ij} = \frac{2(E_i - E_j)}{3}\mathbb{E}_i\left[\frac{r\psi_j^\theta(r)}{\psi_i^\theta(r)}\right]\mathbb{E}_j\left[\frac{r\psi_i^\theta(r)}{\psi_j^\theta(r)}\right] \quad (21)$$

## Nodal surface analysis of wave functions

To visualize the nodal surfaces of benzene, we calculated the wave function values on a two-dimensional slice of a 90-dimensional space. Initially, we fixed a 90-dimensional electron configuration at representative positions for the electronic structure of benzene, effectively shielding 12 electrons near the six carbon atoms due to the effective core approximation. This configuration aligns with the positions determined in the study by Liu et al. [39,46],. For visualization purposes, we applied slight perturbations to this setup.To construct a slice of this 90-dimensional space, we moved one spin-up electron within a two-dimensional square, while keeping the other 29 electrons stationary. Then, for each point on the slice, we applied FermiNet to calculate the wave function values and displayed these values in logarithmic scale. Points with smaller values represent the nodes on each slice. It's important to note that since FermiNet's output is non-normalized, the range of displayed values can vary significantly between different FermiNet diagrams.

### Orbital analysis of the wave function

To understand the shape of each neural network electron orbital constituting the wave function, we fixed the coordinates of all other electrons and moved a specific electron across a three-dimensional spatial grid to calculate its wave function values. Subsequently, we visualized the three-dimensional isosurfaces. By squaring all orbitals of all determinants and then summing them up, we obtain the final electron density. This approach provides a comprehensive view of the spatial distribution of the electron density, essential for understanding the electronic structure of the system under study.